\documentclass[%
 reprint,
 amsmath,amssymb,
 aps,
superscriptaddress, 
]{revtex4-2}
\usepackage{siunitx}
\usepackage{appendix}
\usepackage{graphicx}
\usepackage{dcolumn}
\usepackage{bm}
\usepackage{hyperref}
\usepackage{xcolor}
\usepackage[linesnumbered,ruled,vlined]{algorithm2e}
\usepackage{algorithmic}
\usepackage{listings}
\usepackage{amsmath}

\renewcommand{\eqref}[1]{Eq. (\textup{\ref{#1}})}

\SetKwInput{KwInput}{Input}                

\begin{document}

\title{Accelerating ptychographic reconstructions using spectral initializations}
\author{Lorenzo Valzania}
\thanks{These authors equally contributed to this work. \\
Corresponding author: lorenzo.valzania@lkb.ens.fr}
\affiliation{Laboratoire Kastler Brossel, \'Ecole Normale Sup\'erieure - Paris Sciences et Lettres (PSL) Research University, Sorbonne Universit\'e, Centre National de la Recherche Scientifique (CNRS) UMR 8552, Coll\`ege de France, 24 rue Lhomond, 75005 Paris, France}
\affiliation{Laboratory for Transport at Nanoscale Interfaces, Empa, Swiss Federal Laboratories for Materials Science and Technology, 129 Uberlandstrasse, Dubendorf 8600, Switzerland}
\author{Jonathan Dong}
\thanks{These authors equally contributed to this work. \\
Corresponding author: lorenzo.valzania@lkb.ens.fr}
\affiliation{Laboratoire Kastler Brossel, \'Ecole Normale Sup\'erieure - Paris Sciences et Lettres (PSL) Research University, Sorbonne Universit\'e, Centre National de la Recherche Scientifique (CNRS) UMR 8552, Coll\`ege de France, 24 rue Lhomond, 75005 Paris, France}
\affiliation{Laboratoire de Physique de l'\'Ecole Normale Sup\'erieure - Paris Sciences et Lettres (PSL) Research University, Sorbonne Universit\'e, CNRS, Universit\'e de Paris, 24 rue Lhomond, 75005 Paris, France}
\author{Sylvain Gigan}
\affiliation{Laboratoire Kastler Brossel, \'Ecole Normale Sup\'erieure - Paris Sciences et Lettres (PSL) Research University, Sorbonne Universit\'e, Centre National de la Recherche Scientifique (CNRS) UMR 8552, Coll\`ege de France, 24 rue Lhomond, 75005 Paris, France}

\date{\today}



\begin{abstract}
Ptychography is a promising phase retrieval technique for label-free quantitative phase imaging. Recent advances in phase retrieval algorithms witnessed the development of spectral methods, in order to accelerate gradient descent algorithms. Using spectral initializations on experimental data, for the first time we report three times faster ptychographic reconstructions than with a standard gradient descent algorithm and improved resilience to noise. Coming at no additional computational cost compared to gradient-descent-based algorithms, spectral methods have the potential to be implemented in large-scale iterative ptychographic algorithms.
\end{abstract}

\maketitle



\section{Introduction}

Ptychography is a computational imaging technique that enables label-free, quantitative phase imaging \cite{rodenburg2008ptychography}. It is based on a simple principle: scan a probe across a sample, collect the corresponding intensity diffraction patterns (also known as 'ptychograms'), and reconstruct an image of the object of interest. Because it does not require complex optical elements, it has been adapted to a variety of settings and spectral ranges, from electron microscopy \cite{jiang2018electron}, for which it was originally conceived in the late 60s \cite{hoppe1969diffraction}, to X-rays \cite{pfeiffer2018x} and extreme ultra-violet light \cite{seaberg2014tabletop}, all the way down to the terahertz (THz) range \cite{valzania2018terahertz}.

The computational reconstruction in ptychography requires to solve a phase retrieval problem, where the phase of the diffracted electric field has to be recovered from intensity-only measurements. Such a problem is tractable if a minimum overlap ratio (empirically  estimated around 60\% \cite{bunk2008influence}) is ensured between subsequent probe positions. However, as a non-linear and non-convex optimization problem, it is still not completely understood and convergence towards its global minimum is not guaranteed.

One way to avoid local minima is to provide an initial estimate already close to the solution, and further refine it with iterative algorithms. To this end, spectral methods have recently been proposed for the general phase retrieval problem with independent and identically distributed (i.i.d.) random measurements, solved with a gradient descent (GD) approach \cite{candes2015wirtinger}. The initial estimate is defined as the leading eigenvector of a covariance matrix, constructed from the experimental measurements and the acquisition parameters. The quest for improved reconstruction performance culminated in the derivation of optimal spectral methods for the random measurements setting \cite{mondelli2019fundamental,luo2019optimal}. Due to these breakthroughs, spectral methods for phase retrieval are rather well understood theoretically.

Although spectral initializations to solve ptychography have been showcased on simulated measurements \cite{marchesini2016alternating}, no gain was ever reported when employed on experimental data. Here, we provide the first experimental demonstration of ptychographic reconstructions improved and accelerated by spectral initializations.

\section{Methods}

\begin{figure}[htb]
    \centering\includegraphics[scale = 0.45]{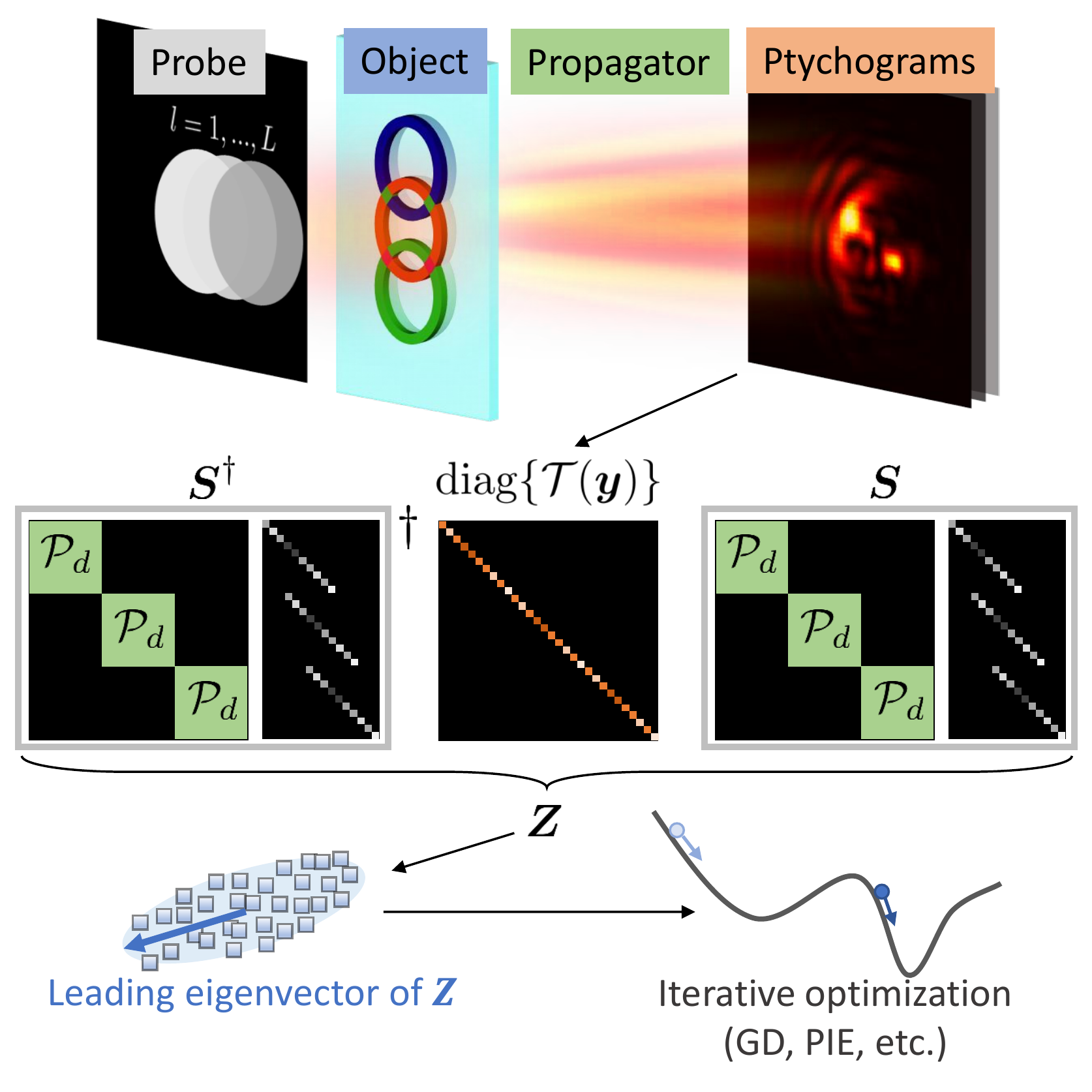}
    \caption{Sketch of a ptychography setup and algorithmic pipeline with our spectral method. Top: an aperture generates the probe $a(x)$, that is scanned with shifts $x^{(l)}$ across an object of interest $\psi(x)$, for $l = 1, \ldots, L$. For each probe position, a camera records a ptychogram $y^{(l)}(x)$ after free space propagation by a distance $d$. Middle: the ptychograms are preprocessed and used to construct a weighted covariance matrix $\boldsymbol{Z}$. Bottom: the spectral estimate, obtained as the leading eigenvector of $\boldsymbol{Z}$, is already close to the solution and is further refined using an iterative optimization (dark blue point). Initial estimates further away from the solution (light blue point) may be stuck in local minima. GD: gradient descent; PIE: ptychographic iterative engine.}
    \label{fig1}
\end{figure}

Let us begin by describing a ptychogram $y^{(l)}(x)$ with the following forward model
\begin{equation}
    \label{eq:ptycho_phys}
    y^{(l)}(x) = |\mathcal{P}_d \{ a(x - x^{(l)}) \,  \psi(x) \}|^2 \, ,
\end{equation}
where $x$ is a two-dimensional spatial coordinate, $\psi(x)$ is the complex transmission function of the object, scanned with the probe $a(x)$ at the positions $x^{(l)}$ for $l = 1, \ldots, L$, with $L$ being the total number of images. $\mathcal{P}_d$ is a known linear operator describing the transmission through the optical system of optical length $d$ between the object plane and the detector plane. We hereby point out that the model equally applies to Fourier ptychography \cite{konda2020fourier}, if one regards $\psi(x)$ and $a(x)$ as the Fourier transform of the object transmission and the probe functions respectively, and $\mathcal{P}_d$ is an inverse Fourier transform. In what follows the case of ptychography will be considered, with a setup sketched in the top row of Fig. \ref{fig1}.

To recover the object, ptychography solves a phase retrieval problem that can be formulated in the vectorized form 
\begin{equation}
    \boldsymbol{y} = |\boldsymbol{S} \boldsymbol{\psi}|^2 \, .
    \label{eq: general phase retrieval}
\end{equation}
Here, $\boldsymbol{\psi} \in \mathbb{C}^n$ represents the vector of the unknown object transmission coefficients at each of its $n$ pixels, $\boldsymbol{S} \in \mathbb{C}^{p \times n}$ is the sensing matrix, with $p$ being the number of measured intensity values (i.e., $p = Lm$, where $m$ is the number of camera pixels), $\boldsymbol{y} \in \mathbb{R}^p$ is the vector collecting the ptychograms, and $|\cdot|$ denotes the element-wise modulus operation.\\
\indent When the entries of $\boldsymbol{S}$ are i.i.d. random variables, $\boldsymbol{\psi}$ is initially estimated as the leading eigenvector of the following $n \times n$ matrix (Fig. \ref{fig1}, middle row):
\begin{equation}
    \boldsymbol{Z} \equiv \boldsymbol{S}^\dag \text{ diag}\{\mathcal{T}(\boldsymbol{y})\} \boldsymbol{S} \, ,
    \label{eq: general spectral method}
\end{equation}
where $\boldsymbol{S}^\dag$ denotes the Hermitian conjugate of the matrix $\boldsymbol{S}$, $\mathcal{T}$ is a preprocessing function acting element-wise on the measurements, and $\text{diag}\{\boldsymbol{u}\}$ denotes a diagonal matrix with a general vector $\boldsymbol{u}$ on its main diagonal.
A widely used method to compute the leading eigenvector is power iterations, that converge exponentially towards the eigenvector corresponding to the largest eigenvalue in absolute value \cite{dongarra1998numerical}. This estimate is finally used as the first estimate of an iterative optimization algorithm like gradient descent or, in the case of ptychography, dedicated routines like the ptychographic iterative engine (PIE) \cite{rodenburg2008ptychography}. The reconstruction pipeline is graphically summarized in the bottom row of Fig. \ref{fig1}.

In order to apply spectral methods to ptychography, the forward model in the continuous domain of \eqref{eq:ptycho_phys} needs to be brought into the discrete vectorized form of \eqref{eq: general phase retrieval}. Let $ \boldsymbol{a}^{(l)} \in \mathbb{C}^{m \times n}$ be the discrete and vectorized form of $a(x-x^{(l)})$, and $\boldsymbol{\mathcal{P}_d} \in \mathbb{C}^{m \times m}$ be the matrix computing the linear transform $\mathcal{P}_d$. Now we can adopt the matrix factorization used in \cite{marchesini2016alternating} and finally write $\boldsymbol{y} = |\boldsymbol{S} \boldsymbol{\psi}|^2$, with $\boldsymbol{S} \equiv \boldsymbol{PA}$ and
\begin{equation}
    \boldsymbol{P} \equiv
        \begin{bmatrix}
            \boldsymbol{\mathcal{P}_d} & \cdots & 0 \\
            \vdots & \ddots & \vdots \\
            0 & \cdots & \boldsymbol{\mathcal{P}_d}
        \end{bmatrix}
        \in \mathbb{C}^{Lm \times Lm} \, , \, \, \boldsymbol{A} \equiv
        \begin{bmatrix}
            \boldsymbol{a}^{(1)} \\
            \vdots \\
            \boldsymbol{a}^{(L)}
        \end{bmatrix}
        \in \mathbb{C}^{Lm \times n} \, ,
\end{equation}
from which, using \eqref{eq: general spectral method} and the unitarity of $\boldsymbol{P}$ (i.e., $\boldsymbol{P}^\dag = \boldsymbol{P}^{-1}$), one obtains
\begin{equation}
    \boldsymbol{Z} = \boldsymbol{A}^\dag \boldsymbol{P}^{-1} \text{ diag}\{\mathcal{T}(\boldsymbol{y})\} \boldsymbol{P} \boldsymbol{A} \, .
\end{equation}
Our choice of the preprocessing function, reading
\begin{equation}
    \mathcal{T}(\boldsymbol{y'}) \equiv \text{max}(-\beta, 1-1/\boldsymbol{y'}) \, ,
    \label{eq: preprocessing}
\end{equation}
where $\boldsymbol{y'}$ collects the ptychograms, each normalized to its own mean intensity, and $0 < \beta < 1$, was inspired by the theoretical works in \cite{luo2019optimal, mondelli2019fundamental, wang2017solving}. For $\boldsymbol{y'} \gtrsim 1$, the function corresponds to the optimal preprocessings of \cite{mondelli2019fundamental, luo2019optimal}. The lower bound set at low intensity values prevents large negative eigenvalues of $\boldsymbol{Z}$, that will degrade the performance of the power iteration algorithm. It is indeed known that power iterations return the eigenvector corresponding to the largest eigenvalue \emph{in absolute value}. In addition, it reminds of the truncated amplitude flow algorithm, whose resilience to noise has been proved \cite{wang2017solving}. After plugging \eqref{eq: preprocessing} into \eqref{eq: general spectral method}, the power method resembles a GD update, proving that no computational complexity is added when replacing one GD iteration with one power iteration (see Appendix \ref{app:gdcomparison}).

In the box Algorithm \ref{algo} we summarize the overall algorithm, consisting of the combination of a spectral initialization and a GD optimization with amplitude loss function $\mathcal{L}(\boldsymbol{\psi}) \equiv \| \sqrt{\boldsymbol{y}} - | \boldsymbol{S} \boldsymbol{\psi} | \|^2_2$, where $\| \cdot \|_2$ indicates the $L^2$-norm.
The corresponding code is available at Ref. \cite{github}, and more details on a memory-efficient implementation of power iterations are outlined in Appendix \ref{app:non-vectorized}.

\begin{algorithm}[htbp]
    \DontPrintSemicolon
    \caption{Solve ptychography with a spectral initialization}
    \label{algo}
    \KwInput{Measurements $\boldsymbol{y}$, sensing matrix $\boldsymbol{S}$, preprocessing function $\mathcal T$, initial estimate $\boldsymbol{\psi}_0$, number of power iterations $M$, number of GD iterations $N$, step size $\gamma$}
    $\boldsymbol{Z} = \boldsymbol{S}^\dag \text{ diag\{} \mathcal T(\boldsymbol{y})\} \boldsymbol{S}$ \\
    \tcc{Spectral initialization}
    \For{$t = 1, \ldots, M$}
    { 
    	$\tilde{\boldsymbol{\psi}}_{t} = \boldsymbol{Z} \boldsymbol{\psi}_{t-1}$ \\
    	$\boldsymbol{\psi}_{t} = \tilde{\boldsymbol{\psi}}_{t} / \| \tilde{\boldsymbol{\psi}}_{t} \|$
    }
    \tcc{Gradient descent}
    \For{$t = M+1, \ldots, M+N$}
    { 
    	$\nabla \mathcal{L}(\boldsymbol \psi) = \boldsymbol{S}^\dag \boldsymbol{S}\boldsymbol \psi_{t-1} - \boldsymbol{S}^\dag \text{ diag}\{ \sqrt{\boldsymbol y} / |\boldsymbol{S} \boldsymbol \psi_{t-1}| \} \boldsymbol{S} \boldsymbol \psi_{t-1}$ \\
    	$\boldsymbol{\psi}_{t} = \boldsymbol{\psi}_{t-1} - \gamma \nabla \mathcal{L}(\boldsymbol \psi)$
    }
    \textbf{return} $\boldsymbol{\psi}_{M+N}$
\end{algorithm}

\section{Experiments}

Experiments were performed using the THz imaging setup at Empa, the Swiss Federal Laboratories for Materials Science and Technology, equipped with a far-infrared gas laser (FIRL 100, Edinburgh Instruments, Livingston, Scotland) emitting several tens of mW of continuous-wave power at the wavelength $\lambda$ = 96.5 $\si{\micro\meter}$. An uncooled microbolometer array detector featuring $m$ = 480 $\times$ 640 pixels on a pitch of 17 $\si{\micro\meter}$ (Gobi-640-GigE, Xenics, Leuven, Belgium), was used as a THz camera \cite{hack2016comparison}. The object was a 2-mm-thick polypropylene (PP, refractive index $n_{PP}$ = 1.51 and absorption coefficient $\alpha_{PP}$ = 1.5 cm$^{-1}$ at $\lambda$ = 96.5 $\si{\micro\meter}$ \cite{lee2009principles}) slab where three intersecting rings were engraved by laser ablation at depths in the range 37-234 $\si{\micro\meter}$. Fig. \ref{fig1} shows the wrapped phase shift distribution induced by the object at $\lambda$ = 96.5 $\si{\micro\meter}$. We let a plane wave diffract through a circular aperture with a diameter of 3 mm and propagate by 12 mm before impinging on the object. The object was scanned across a grid of 21 $\times$ 7 points at overlap ratios of 87\% and 80\% \cite{bunk2008influence} along its vertical and horizontal axis in Fig. \ref{fig1}, respectively. The ptychograms were recorded after a free space propagation of $d = \SI{9.1}{\milli\meter}$. This resulted in a Fresnel number around 2, making the angular spectrum propagator a suitable choice for computing $\mathcal{P}_d$ \cite{valzania2018terahertz, goodman2005introduction}.

\section{Results}

\begin{figure}[htbp]
    \centering\includegraphics[scale = 0.72]{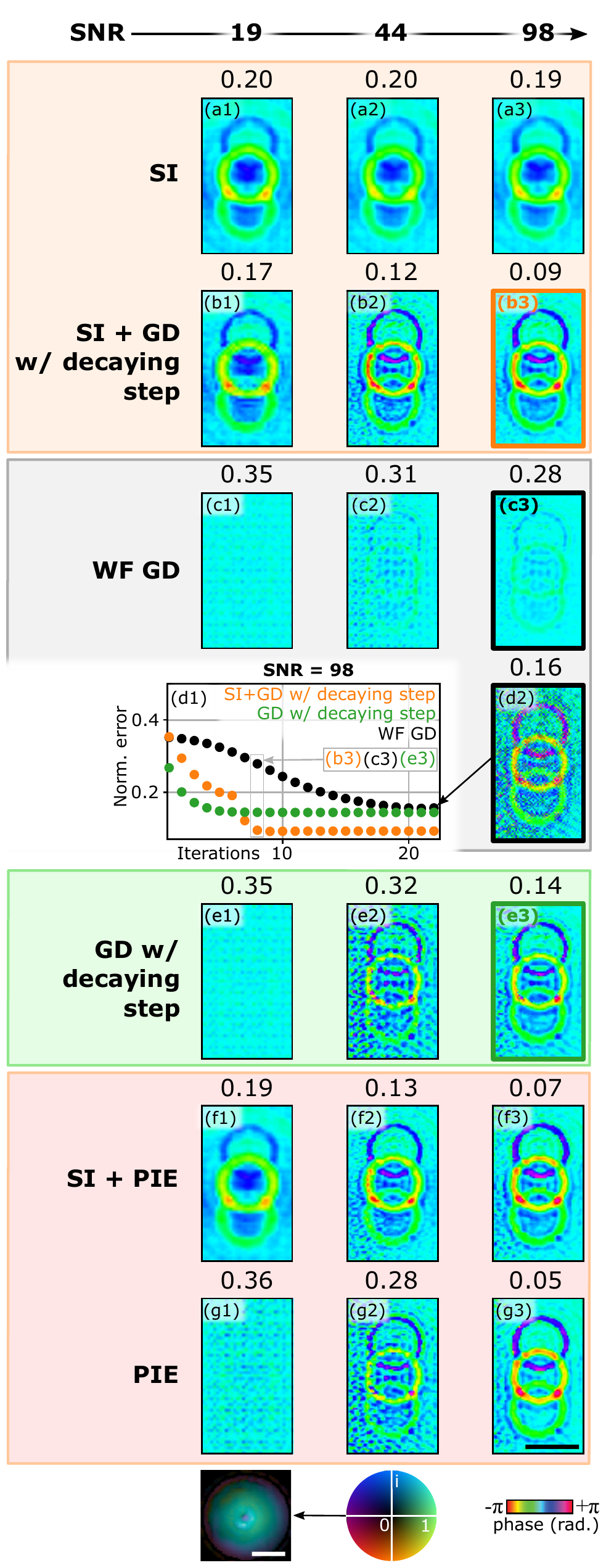}
    \caption{Reconstructions of the phase object in Fig. \ref{fig1} with different algorithms (rows (a)-(c) and (e)-(g)) at increasing SNR (labels (1)-(3)). In (d1) the reconstruction curves for three algorithms at the highest SNR are plotted, while (d2) shows the reconstruction from WF GD. A normalized error metric \cite{maiden2009improved} is given above each reconstruction, where all reconstructions are compared with a reference reconstruction from an rPIE algorithm, capable of retrieving both the object and the probe \cite{maiden2017further}. SI: spectral initialization; GD: gradient descent; WF GD: Wirtinger flow GD; PIE: ptychographic iterative engine. Bottom row, left to right: probe reconstruction, its two-dimensional colorbar, and the phase colorbar used for the reconstructions of the object. Scale bars: 2 mm.}
    \label{fig2}
\end{figure}

Although spectral methods in combination with preprocessing functions similar to \eqref{eq: preprocessing} are well grounded in the theory of random sensing, their effectiveness and resilience to noise are also maintained when $\boldsymbol{S}$ is not random to a certain extent, as in our experimental setting. To prove this, we analyzed three datasets, collected at different signal-to-noise ratios (SNR) by duly tuning the laser power and denoised with the procedure borrowed from \cite{williams2010fresnel}. Results for each noise level are compiled in the columns of Fig. \ref{fig2}, labelled with their corresponding SNR, calculated as the ratio between the mean intensity in the center of each ptychogram and that at its outermost pixels, where the diffracted intensity is negligible. Starting from a flat estimate in both amplitude and phase, the power method yielded the spectral estimates in row (a) (the normalized reconstruction error \cite{maiden2009improved, maiden2017further} is given above each image). Although we cannot guarantee that the leading eigenvector of $\boldsymbol{Z}$ is the sought solution even in the absence of randomness, we have indeed observed a decrease of the loss $\mathcal{L}$ in the early iterations of the power method, which can be explained by comparing one iteration of the power method with one GD iteration (see Appendix \ref{app:gdcomparison}). When such a decrease came to a halt, power iterations were stopped, and the obtained spectral estimate was used as the initial guess for GD, run with an exponentially decreasing step size, compatible with a backtracking line search, to ensure stable convergence. At all noise levels, two GD iterations are enough to provide a satisfactory reconstruction of the object, shown in row (b). For comparison, we also performed reconstructions with a standard Wirtinger flow GD algorithm without spectral initialization, using the same initial estimate provided to the power method in row (a) and the step size recommended in \cite{candes2015wirtinger}. The results, after the same total number of iterations run in row (b) (summing power iterations and GD iterations), are shown in row (c). Note that in the noisiest case, the reconstruction is dominated by high frequency artifacts at the periodicity of the acquisition scan, caused by a signal bias in the measurements \cite{thurman2009phase}. With the highest SNR convergence is reached, however at a three times higher number of iterations, as can be observed comparing the orange and black curves in Fig. \ref{fig2}(d1), and at a worse final reconstruction (Fig. \ref{fig2}(d2)). The reconstruction without spectral initialization can be improved and accelerated with an exponentially decaying step size (see the green curve in Fig. \ref{fig2}(d1) for the highest SNR case and row (e) for all the noise levels). However, its quality does not reach that of the reconstructions benefiting from the spectral initialization. Notably, an error reduction by a factor of 2 can be observed at the lowest SNR.

Although conceived in combination with GD optimizations, a spectral estimate may be used as initial guess for any iterative ptychographic algorithm. Rows (f) and (g) show the corresponding reconstructions after a PIE algorithm initialized with the spectral estimate and with a flat distribution, respectively.
Better reconstructions are obtained with PIE than with GD in the case of highest SNR, as already demonstrated in \cite{hawkes2019springer} and owing to the spatially varying step size \cite{guizar2008phase}. The advantage brought by the spectral estimate is clear in the two cases with lowest SNR, whereas with the cleanest data PIE alone appears to just win out, judging from the reconstruction errors (although the visual difference is more subtle).

All the results shown above were obtained after reconstructing the probe $a(x)$ (plotted in the last row of Fig. \ref{fig2} with a two-dimensional colormap encoding its amplitude in the lightness and its phase in the hue) from the highest SNR dataset with a regularized PIE (rPIE) algorithm \cite{maiden2017further}.
Nevertheless, spectral initializations may be implemented in any self-calibrating algorithm which jointly optimizes for both the object and the acquisition parameters. When performing algorithmic self-calibration with alternating projections, the unknowns are optimized one at a time, while leaving the others constant \cite{chen2018quantitative}. In this perspective, spectral methods would naturally have a place as the object update step at the earliest iterations of self-calibrating large-scale algorithms using alternating projections. They would efficiently drive the solution beyond local minima induced by experimental noise. This procedure would then be followed by the refinement of the acquisition parameters.

The matrix multiplication involved in the power method implies that the solution is updated after all the ptychograms have been used once. This makes our spectral method a 'batch' algorithm, like gradient-descent-based pytchographic solvers \cite{thibault2012maximum}. On the other hand all the variants of the PIE \cite{maiden2017further} are 'serial' algorithms, delivering a new estimate of the solution every time a ptychogram is used. While the former display a better resilience to noise and enjoy parallel computing, the latter typically converge faster \cite{yeh2015experimental, guizar2008phase, hawkes2019springer}. In order to combine the benefits of both classes, a 'stochastic' method based on 'mini-batches' of the full set of ptychograms has been put forward \cite{odstrvcil2018iterative} and shown to be particularly well suited to the reconstruction of low spatial frequency phases \cite{odstrvcil2019towards}. We hereby point out that the power method is fully compatible with this procedure, as already envisioned in the seminal paper \cite{candes2015wirtinger}.

Finally, note that spectral methods are essentially different from commonly used initialization procedures for iterative ptychographic algorithms. Those are based on synthesizing a low-resolution object from a subset of the measurement data with additional dedicated procedures \cite{tian2014multiplexed}. Thanks to the similarity between GD iterations and power iterations, we could instead \emph{replace} early GD iterations with the spectral method, adding no computational burden to the reconstruction framework while, at the same time, leveraging the full dataset to compute the spectral estimate.

In conclusion, we have accelerated and improved ptychographic reconstructions by spectral methods on experimental data. Our results confirm that their application can go beyond random settings, where they were originally developed. Compared to a reference GD algorithm using no spectral methods, three times faster reconstructions were obtained. At different levels of SNR, reconstructions employing spectral methods outperformed standard reconstructions, reaching a 2-fold reduction of the reconstruction error at low SNR. Although our proof-of-principle reconstructions used the simplest form of optimization algorithms for ptychography, we have envisioned the implementation of spectral methods in large-scale optimization frameworks too.
This would come at no additional computational cost compared to a GD algorithm.

\section*{Funding}
Swiss National Science Foundation (200021\textunderscore 160078/1 and P2BEP2\textunderscore 188152); H2020 European Research Council (SMARTIES-724473).

\section*{Acknowledgments}
LV wishes to thank R. Br{\"o}nnimann, E. Hack and P. Zolliker (Empa) for fabricating the sample and for contributing constructive comments during the data acquisition.

\appendix

\section{Intuition of the spectral estimate}
\vspace{-2mm}
Spectral methods are simple-to-code algorithms to find initial estimates of the solution to phase retrieval problems, usually combined with gradient descent to refine the spectral estimate. The intuition behind these spectral methods is the following: for one intensity measurement $y_i = |\boldsymbol{s}_i^\dagger \boldsymbol{\psi}|^2$ (with $\boldsymbol{s}_i^\dagger$ denoting a row of the matrix $\boldsymbol{S}$), when $\boldsymbol{s}_i$ is correlated with $\boldsymbol{\psi}$ then the measured intensity $y_i$ is increased. On the other hand, if $\boldsymbol{s}_i$ is orthogonal to $\boldsymbol{\psi}$, the measured intensity is equal to 0. Therefore we aim to create a matrix dominated by the sampling vectors aligned with $\boldsymbol{\psi}$. This leads to the definition of a weighted covariance matrix to exploit this information:
\begin{equation}
    \boldsymbol{Z} = \sum_{i=1}^p \mathcal{T}(y_i) \boldsymbol{s}_i \boldsymbol{s}_i^\dagger \,,
\end{equation}
which corresponds to Eq. (3) in the main manuscript. This matrix is essentially a sum of rank-1 matrices $\boldsymbol{s}_i \boldsymbol{s}_i^\dagger$, each weighted according to the measured intensity $y_i$. Such matrices first appeared without preprocessing function $\mathcal{T}$, but it was quickly realized how this operation could improve the performance of spectral methods. In the end, $\mathcal{T}$ only needs to be an increasing function in order to respect the intuition detailed above. 
There is a relatively good understanding of spectral methods in the random setting. As the number of measurements $p$ increases, the leading eigenvector of $\boldsymbol{Z}$ converges towards the sought solution $\boldsymbol{\psi}$. The sample complexity, i.e. how many more measurements $p$ one needs compared to the dimensionality of the problem $n$, is well characterized when $\mathcal{T}$ is bounded above \cite{lu2020phase}. In particular, optimal spectral methods actually outperform gradient descent alone, as they require fewer measurements to provide a meaningful estimate \cite{mondelli2019fundamental}. 

Because of this previous heuristic, the recent theoretical advances of spectral methods may be applied to other non-random settings, beyond ptychography. It would be interesting to investigate whether they may apply to other phase retrieval problems, even for example 3D reconstructions. 

\section{Comparison with GD update}
\vspace{-2mm}
\label{app:gdcomparison}

We would like to emphasize how one power iteration of the spectral method relates to one gradient descent iteration. In particular, as made explicit in the formulas below, the computational complexity of one power iteration is comparable to a gradient descent iteration. 

Let us start with the gradient descent update with the amplitude loss function as described in the main text. At each iteration $t > 0$, the current estimate is computed according to the following equation \cite{yeh2015experimental}:
\begin{equation}
    \boldsymbol{\psi}_{t} = \boldsymbol{\psi}_{t-1} + \gamma \left( \boldsymbol{S}^\dag \text{ diag}\{ \sqrt{\boldsymbol{y}} / |\boldsymbol{S} \boldsymbol{\psi}_{t-1}| \} \boldsymbol{S} \boldsymbol{\psi}_{t-1} - \boldsymbol{S}^\dag \boldsymbol{S}\boldsymbol{\psi}_{t-1}\right)
\end{equation}

On the other hand, with the optimal preprocessing function $\mathcal{T}^*(\boldsymbol{y}) = 1 - \langle \boldsymbol{y} \rangle / \boldsymbol{y}$, by performing a linear Taylor expansion in $\sqrt{\boldsymbol{y}/\langle \boldsymbol{y} \rangle}$ near 1, we obtain the preprocessing function $\mathcal{T}(\boldsymbol{y}) = 2 \left( \sqrt{\boldsymbol{y}/\langle \boldsymbol{y} \rangle} - 1\right)$. A power iteration in this case is described by the following map:
\begin{equation}
    \boldsymbol{\psi}_{t} = 2 \left( \boldsymbol{S}^\dag \text{ diag}\left\{ \sqrt{\boldsymbol{y}/\langle \boldsymbol{y}\rangle} \right\} \boldsymbol{S} \boldsymbol{\psi}_{t-1} - \boldsymbol{S}^\dag \boldsymbol{S}\boldsymbol{\psi}_{t-1} \right)
\end{equation}
Additionally, it is possible to add a term $\boldsymbol{\psi}_{t-1}$ by replacing $\boldsymbol{Z}$ by $\boldsymbol{Z} + \boldsymbol{I}$, an operation which does not change the leading eigenvector, but averages the current estimate of the leading eigenvector with the previous one. We thus observe that the two updates are quite similar. Despite the similarities in the equations, note that spectral methods are guaranteed to converge towards the eigenvector of the largest eigenvalue, whereas gradient descent may converge to a local minimum of this non-convex optimization problem. 

\section{Non-vectorized version of power iterations}
\vspace{-2mm}
\label{app:non-vectorized}

The implementation of the matricial formalism involved after the definition of $\boldsymbol{S}$ typically becomes prohibitively large in a usual ptychographic setting. For example, $\boldsymbol P$ is of size $Lm \times Lm$, with $Lm$ the total number of measured pixels, typically much greater than $10^4$. Although this matricial formalism is important to link ptychography with other theoretical settings, it would be unpractical to store $\boldsymbol P$ in memory and repeatedly compute multiplications by $\boldsymbol{P}$ to retrieve the leading eigenvector of the associated matrix $\boldsymbol{Z}$. 

Instead, we perform each power iteration by observing that {a} multiplication by $\boldsymbol P \boldsymbol A$ corresponds to {applying} the ptychographic forward model, a multiplication by $\text{diag}\{\mathcal{T}(\boldsymbol{y})\}$  {corresponds} to an element-wise multiplication with the measured intensities, and {applying} $\boldsymbol A^\dag \boldsymbol P^{-1}$ {is equivalent} to a backward pass in the ptychographic model (i.e. the backpropagation step of gradient-descent methods). 
In practice, this operation can be done sequentially, for each ptychogram indexed by $l = 1, \ldots, L$. Each step provides us with a partial estimate of the solution $\psi_{t+1}^{(l)}(x)$, which is restricted to the region illuminated at the scan position $x^{(l)}$. Using the symbols employed for the continuous-domain model in the main manuscript, we obtain:
\begin{equation}
    \psi_{t+1}^{(l)}(x) = \bar{a}(x-x^{(l)}) \, \mathcal{P}_{-d} \big\{ \mathcal{T} (y^{(l)}(x)) \, \mathcal{P}_{d} \{ a(x-x^{(l)}) \psi_{t}(x)\} \big\} \, ,
\end{equation}
where $\bar{a}(x-x^{(l)})$ denotes the complex conjugate of $a(x-x^{(l)})$. Stitching these partial estimates yields the estimate $\psi_{t+1}(x)$ at the $(t+1)$-th iteration.

\section{Different preprocessing functions}
\vspace{-2mm}




Over the years, spectral methods based on different preprocessing functions have been proposed, e.g. $\mathcal{T}_1$ in \cite{candes2015wirtinger}, $\mathcal{T}_2$ in \cite{marchesini2016alternating}, and the optimal preprocessing function in the noiseless setting $\mathcal{T}^*$ in \cite{luo2019optimal}, defined as:
\begin{equation}
\mathcal{T}_1(\boldsymbol{y}) \equiv \boldsymbol{y} \quad \quad 
\mathcal{T}_2(\boldsymbol{y}) \equiv 
    \begin{cases}
        0 \quad \text{if } \boldsymbol{y} \leq T \\
        1 \quad \text{if } \boldsymbol{y} > T
    \end{cases} \quad
\mathcal{T}^*(\boldsymbol{y}) \equiv 1 - \frac{1}{\boldsymbol{y}} \, ,
\end{equation}

where $T$ is a predefined threshold usually set to a quantile of the intensity distribution (for example keeping the top 20\% values). 
The optimal preprocessing function $\mathcal{T}^*$ was used to obtain the results presented in the main text. Since $\mathcal{T}^*(\boldsymbol{y}) = 1 - \boldsymbol{y}^{-1}$ is not bounded below, some eigenvalues of $\boldsymbol{Z}$ may be negative which is detrimental for power iterations. Although we aim at finding the eigenvector associated to the largest eigenvalue of $\boldsymbol{Z}$, the power method selects the eigenvector corresponding to the largest eigenvalue \emph{in modulus}. Therefore, care must be taken to avoid large negative eigenvalues. To prevent this, in our reconstructions we set a lower bound $-\beta$ to the preprocessed intensities. 

\begin{figure}[htbp]
    \centering\includegraphics[scale = 0.26]{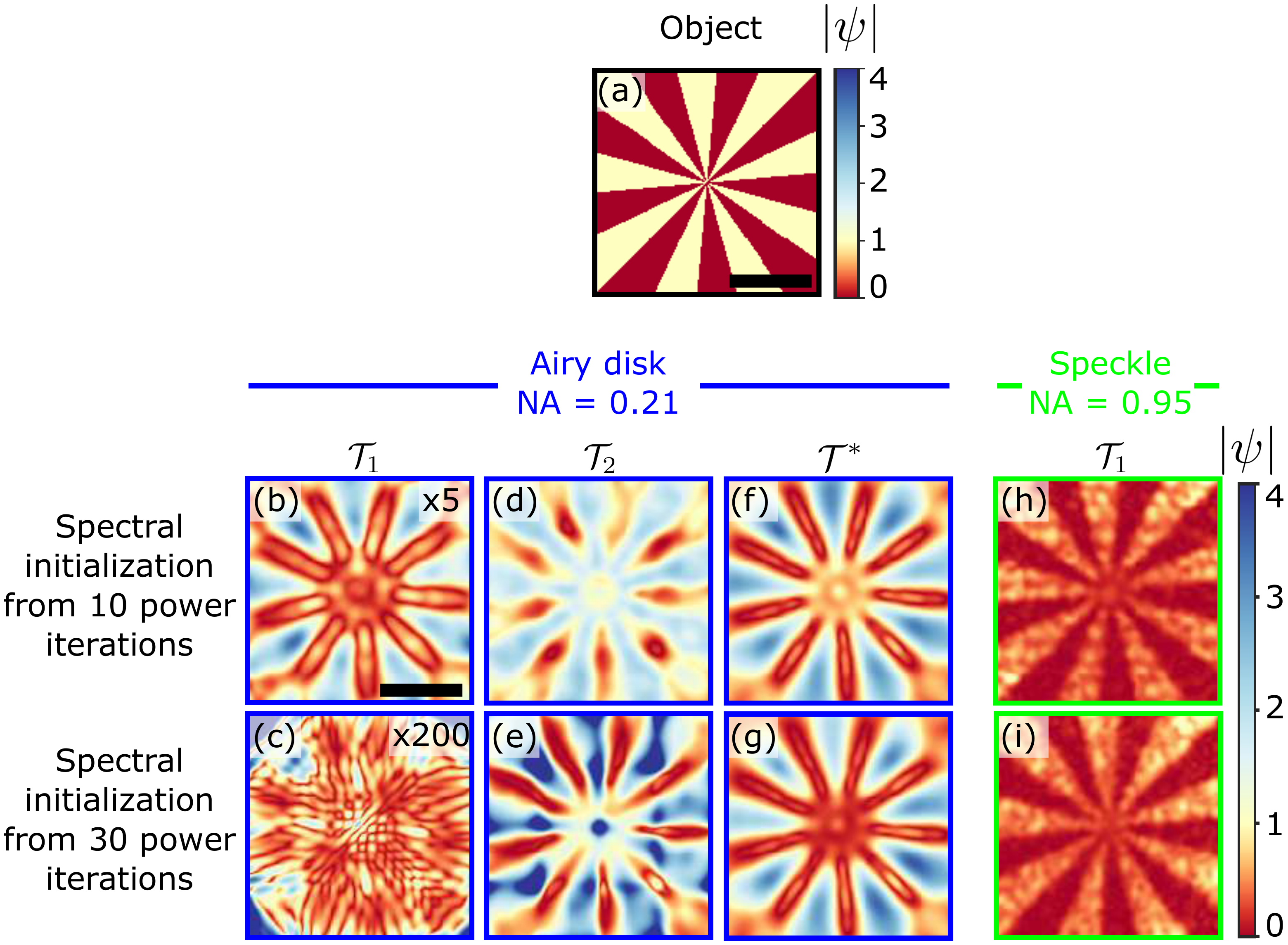}
    \caption{Spectral initializations at different numbers of power iterations, preprocessing and illumination functions (simulated data). Note that, in order for all the images to share the same colorbar, shown on the right, (b) and (c) have been multiplied by 5 and 200, respectively. Scale bar: 1 mm ($\sim$ 10 times the wavelength).}
    \label{fig3}
\end{figure}

To study in more detail the different spectral methods, we performed a numerical experiment with a Siemens star amplitude-only object (Fig. \ref{fig3}(a)), {using an angular spectrum propagator at a Fresnel number of 3.9, a ptychographic overlap ratio of 77\% and} starting the power method from a random initial guess. Fig. \ref{fig3} shows different initialization estimates for various choices of the preprocessing function $\mathcal{T}$, the number of power iterations $M$, and the spatial distribution of the illumination function $a(x)$. We have seen empirically that these three parameters are impacting the most the spectral method performance. Note that when the illumination function is a weakly structured Airy disk (Figs. \ref{fig3}(a-f)), the hypotheses of random sampling required by the spectral method \cite{candes2015wirtinger} break down. For this reason, even in a noiseless setting like the one presented in Fig. \ref{fig3}, we cannot expect to indefinitely approach the solution using power iterations, as confirmed by Figs. \ref{fig3}(b, c). Moreover, the optimality of $\mathcal{T}^*$ is not established with our non i.i.d. random matrix $\boldsymbol S$. However, the preprocessing functions $\mathcal{T}_2$ and $\mathcal{T}^*$ make the estimates more robust to the power iterations, while delivering informative estimates already after 10 power iterations. 

In Figs. \ref{fig3}(h, i), we simulated a random beam with a speckle grain size about 10 times smaller than the illumination shifts, so to boost the diversity of the ptychograms upon translation. This makes the acquisition closer to that of coded diffraction imaging, for which spectral methods were originally developed and have already been applied \cite{candes2015wirtinger, mondelli2019fundamental}. As a result, a much more informative spectral initialization is obtained, with no need to preprocess the measurement data. Besides an increase of the spatial resolution, we notice a more reliable quantification of the amplitude.

These results can also be seen in the context of ptychography with randomized and structured illuminations, which has been implemented with three main advantages: adding diversity in the ptychograms \cite{odstrvcil2019towards, stockmar2013near}, accessing higher spatial frequencies \cite{bian2019ptychographic} as in coherent structured illumination microscopy \cite{wicker2014resolving}, and reducing dynamic range requirements in X-ray imaging \cite{morrison2018x, maiden2013soft}.


\section{Amplitude correction for the spectral initialization}
\vspace{-2mm}

Because the spectral initialization is calculated from an eigenvalue problem, one needs to design a strategy to choose its norm, especially for amplitude objects such as the Siemens star (see previous section). Moreover, this normalization is performed pixel by pixel to account for the fact that all the pixels are not sampled uniformly in ptychography.

In order to obtain physically meaningful values of the modulus of the solution, the spectral initialization after $M$ power iterations $\boldsymbol{\psi}_{M}$ was normalized with the spectral initialization $\boldsymbol{\psi}^{\text{ref}}_{M}$ of a "reference" experiment without object. In other words, $\boldsymbol{\psi}^{\text{ref}}_{M}$ is the leading eigenvector of
\begin{equation}
    \boldsymbol{Z^{\text{ref}}} = \boldsymbol{S}^\dag \text{diag} \{ \mathcal{T}(\boldsymbol{y^{\text{ref}}}) \} \boldsymbol{S} \, , \quad \quad \boldsymbol{y^{\text{ref}}} = |\boldsymbol{S} \boldsymbol{\psi}^{\text{ref}}|^2 \, ,
\end{equation}
with $\boldsymbol{\psi}^{\text{ref}} \equiv [1, \dots, 1]^T \in \mathbb{R}^{n}$ denoting the transmission function through free space and the superscript $^T$ indicating the transpose operation.

\bibliography{biblio}

\end{document}